\documentclass[12pt,prd,aps,amssymb,amsmath,tightenlines,showpacs]{revtex4}
\usepackage{graphicx}
\begin{document}

\def\half{\textstyle{\frac{1}{2}}}
\def\cP{\mathcal P}
\def\cC{\mathcal C}
\def\cT{\mathcal T}

\topmargin=0.0cm

\title{Bound states of $\cP\cT$-symmetric separable potentials}

\author{Carl~M.~Bender${}^1$}
\email{cmb@wustl.edu}

\author{Hugh~F.~Jones${}^2$}\email{h.f.jones@imperial.ac.uk}

\affiliation{${}^1$Department of Physics, Washington University, St. Louis
MO 63130, USA\\
${}^2$Blackett Laboratory, Imperial College, London SW7 2AZ, UK}

\begin{abstract}
All of the $\cP\cT$-symmetric potentials that have been studied so far have been
local. In this paper nonlocal $\cP\cT$-symmetric separable potentials of the
form $V(x,y)=i\epsilon[U(x)U(y)-U(-x)U(-y)]$, where $U(x)$ is real, are
examined. Two specific models are examined. In each case it is shown that there
is a parametric region of the coupling strength $\epsilon$ for which the $\cP
\cT$ symmetry of the Hamiltonian is unbroken and the bound-state energies are
real. The critical values of $\epsilon$ that bound this region are calculated.
\end{abstract}
\pacs{11.30.Er, 02.30.Em, 03.65.-w}
\maketitle

\section{Introduction}
\label{s1}
The time-independent Schr\"odinger equation $H\psi=E\psi$, when written in
coordinate space, takes the form of a continuous matrix eigenvalue
problem
\begin{equation}
\int dy\,H(x,y)\psi(y)=E\psi(x).
\label{eq1}
\end{equation}
However, all previous studies of non-Hermitian $\cP\cT$-symmetric
quantum-mechanical Hamiltonians have focused on Hamiltonians that in the
coordinate representation are diagonal and symmetric. Such Hamiltonians take
the form $H=p^2+V(x)$, where the condition of $\cP\cT$ symmetry is that $V^*(x)
=V(-x)$. When expressed as a matrix, $H$ is clearly diagonal,
\begin{equation}
H(x,y)=\partial_x\partial_y\delta(x-y)+V(x)\delta(x-y),
\label{eq2}
\end{equation}
and for Hamiltonians of this form the Schr\"odinger eigenvalue problem
(\ref{eq1}) is the differential equation
\begin{equation}
-\psi''(x)+v(x)\psi(x)=E\psi(x).
\label{eq3}
\end{equation}

The first $\cP\cT$-symmetric Hamiltonians that were examined in detail belong to
the class of diagonal Hamiltonians \cite{R1,R2,R3,R4}
\begin{equation}
H=p^2+x^2(ix)^\epsilon\qquad(\epsilon\geq0).
\label{eq4}
\end{equation}
The Hamiltonians (\ref{eq4}) can have many different spectra depending on the
large-$|x|$ boundary conditions that are imposed on the solutions to the
corresponding time-independent Schr\"odinger eigenvalue equation
\begin{equation}
-\psi''(x)+x^2(ix)^\epsilon\psi(x)=E\psi(x).
\label{eq5}
\end{equation}
The boundary conditions on $\psi(x)$ are imposed in Stokes' wedges in the
complex-$x$ plane. At the edges of the Stokes' wedges both linearly independent
solutions to (\ref{eq5}) are oscillatory as $|x|\to\infty$. However, in the
interior of the wedges one solution decays exponentially and the linearly
independent solution grows exponentially. The eigenvalues $E$ are determined by
requiring that $\psi(x)$ decay exponentially in two nonadjacent wedges.
Ordinarily, the eigenvalues are complex, but if the two wedges are $\cP
\cT$-symmetric relative to the imaginary-$x$ axis, then all the eigenvalues are
real. (The $\cP\cT$ reflection of the complex number $x$ is the number $-x^*$.)

We emphasize that a Hamiltonian need not be diagonal. Of course, for nondiagonal
Hamiltonians it is difficult to solve the Schr\"odinger eigenvalue problem
because it takes the form of the Volterra intego-differential equation
(\ref{eq1}) rather than the differential equation (\ref{eq3}). However, there is
an interesting special solvable class of nondiagonal Hamiltonians for which the
potential is separable; these Hamiltonians have the form
\begin{equation}
H(x,y)=\partial_x\partial_y\delta(x-y)+U(x)U(y).
\label{eq6}
\end{equation}
Potentials of this form are discussed in Ref.~\cite{R5} and are interesting
because they can be used in studies of scattering processes.

The purpose of this paper is to consider the case of complex $\cP\cT$-symmetric
separable potentials. We show that such potentials can have unbroken (and
broken) $\cP\cT$-symmetric regions in which the bound states are real (and
complex). In Sec.~\ref{s2} we show how to find bound states for real Hermitian
separable potentials and in Secs.~\ref{s3} and \ref{s4} we show how to find
bound states for complex $\cP\cT$-symmetric potentials. Finally, in
Sec.~\ref{s5} we make some brief concluding remarks.

\section{Bound states of Hermitian separable potentials}
\label{s2}

Let us consider the case of a Hamiltonian of the form $H=p^2+V(x,y)$, where
$V(x,y)=gU(x)U(y)$ is a separable potential and $g$ is a coupling strength. Note
that if $U(x)$ is real, then $H$ is Hermitian. The Schr\"odinger eigenvalue
equation for this Hamiltonian is
\begin{equation}
-\psi''(x)+gU(x)\int_{-\infty}^\infty dy\,\psi(y)U(y)=E\psi(x),
\label{eq7}
\end{equation}
which is a linear integro-differential equation in Volterra form.

In general, the potential $V(x,y)$, may have bound states and scattering states.
These states are distinguished by their large-$|x|$ behavior. For a scattering
state the wave function $\psi(x)$ does not vanish as $|x|\to\infty$, but for a
bound state $\psi(x)\to0$ as $|x|\to\infty$. Furthermore, if $\psi(x)$ vanishes
for large $|x|$, then (\ref{eq7}) implies that $U(x)$ must also vanish. Thus, if
we wish to solve (\ref{eq7}) for a bound state, it is valid to perform a Fourier
transform:
\begin{equation}
(E-p^2)\tilde\psi(p)=g\tilde U(p)\int_{-\infty}^\infty dy\,\psi(y)U(y),
\label{eq8}
\end{equation}
where the Fourier transform of $\psi(x)$ is given by
\begin{equation}
\tilde\psi(p)\equiv\int_{-\infty}^\infty dx\,e^{ipx}\psi(x)
\label{eq9}
\end{equation}
and the inverse Fourier transform of $\tilde\psi(p)$ is given by
\begin{equation}
\psi(x)\equiv\frac{1}{2\pi}\int_{-\infty}^\infty dp\,e^{-ipx}\tilde\psi(p).
\label{eq10}
\end{equation}
Note also that the integral in (\ref{eq8}) can be written as a convolution of
two Fourier transforms:
\begin{equation}
\int_{-\infty}^\infty dy\,\psi(y)U(y)=\frac{1}{2\pi}\int_{-\infty}^\infty dq\,
\tilde U(-q)\tilde\psi(q).
\label{eq11}
\end{equation}

To proceed, we multiply (\ref{eq8}) by $\tilde U(-p)=\tilde U^*(p)$, integrate
with respect to $p$, and assume that
\begin{equation}
\alpha=\int_{-\infty}^\infty dq\,\tilde U(-q)\tilde\psi(q)\neq0.
\label{eq12}
\end{equation}
The result is the secular eigenvalue condition
\begin{equation}
1=\frac{g}{2\pi}\int_{-\infty}^\infty dp\,\frac{|\tilde U(p)|^2}{E-p^2}.
\label{eq13}
\end{equation}
The integral in this equation must exist, and it clearly exists if $E$ has a
nonzero imaginary part. However, complex energy is associated with scattering,
and we are not concerned with scattering states in this paper.

There are two possible ways to have real-energy bound-state solutions to
(\ref{eq13}). First, $E$ may be negative. In this case the integral exists but
it is negative, and this requires that the coupling constant $g$ be negative.
While this possibility is viable for the case of Hermitian Hamiltonians, it is
not viable for the case of non-Hermitian $\cP\cT$-symmetric separable
Hamiltonians, as we will show in Sec.~\ref{s2}. Another way to have a bound
state is for $\tilde U(p)$ to vanish when $p^2=E$, where $E$ is a number to be
determined consistently, and if this happens, the integral in (\ref{eq13}) will
exist and there can be a bound state.

Let us construct a potential for which we can solve the one-bound-state problem
analytically: We assume for simplicity that $\tilde U(p)$ has the form
\begin{equation}
\tilde U(p)=e^{-p^2/2}(E-p^2).
\label{eq14}
\end{equation}
Then, from evaluating the integral in (\ref{eq13}) we get an equation for $E$ in
terms of $g$:
\begin{equation}
E=\frac{1}{2}+\frac{2\sqrt{\pi}}{g}.
\label{eq15}
\end{equation}
We also obtain the result that
\begin{equation}
U(x)=\frac{1}{\sqrt{2\pi}}\left(\frac{2\sqrt{\pi}}{g}-\frac{1}{2}+x^2\right)
e^{-x^2/2}.
\label{eq16}
\end{equation}

We now verify the consistency of this calculation by showing that $\alpha$ in
(\ref{eq12}) is nonzero. To do so, we calculate $\tilde\psi(p)$ from (\ref{eq8})
and find that $\tilde\psi(p)=g\alpha e^{-p^2/2}/(2\pi)$, and thus that $\psi(x)=
g\alpha(2\pi)^{-3/2}e^{-x^2/2}$. We then evaluate (\ref{eq12}) and find that it
reduces to the identity $\alpha=\alpha$. Thus, $\alpha$ is an arbitrary
normalization constant and the bound-state solution is internally consistent.

\section{Bound states of $\cP\cT$-symmetric separable potentials}
\label{s3}

For the non-Hermitian case we assume that the potential has the $\cP
\cT$-symmetric form $V(x,y)=\epsilon i[W(x,y)-W(-x,-y)]$, where $W(x,y)$ is
real. We then further specialize the potential by making it separable and
symmetric under the interchange of $x$ and $y$: $V(x,y)=i\epsilon[U(x)U(y)-U(-x)
U(-y)]$, where $U(x)$ is real. This gives the Schr\"odinger equation
\begin{equation}
-\psi''(x)+i\epsilon\left[U(x)\int_{-\infty}^\infty dy\,\psi(y)U(y)
-U(-x)\int_{-\infty}^\infty dy\,\psi(y)U(-y)\right]=E\psi(x).
\label{eq17}
\end{equation}
We can rewrite this equation as
\begin{equation}
-\psi''(x)+\frac{i\epsilon}{2\pi}\left[\alpha U(x) -\beta U(-x)\right]=E\psi(x),
\label{eq18}
\end{equation}
where $\alpha$ and $\beta$ are expressed as convolutions of the Fourier
transform of $U$ and the Fourier transform of $\psi$:
\begin{equation}
\alpha=\int_{-\infty}^\infty dq\,\tilde U(-q)\tilde\psi(q),\qquad
\beta=\int_{-\infty}^\infty dq\,\tilde U(q)\tilde\psi(q).
\label{eq19}
\end{equation}
We assume that $\alpha$ and $\beta$ are nonzero.

As in the Hermitian case discussed in Sec.~\ref{s2}, we solve the
Schr\"odinger equation (\ref{eq18}) by taking a Fourier transform:
\begin{equation}
\tilde\psi(p)=\frac{i\epsilon}{2\pi}\left[\frac{\tilde U(p)}{E-p^2}\alpha
-\frac{\tilde U(-p)}{E-p^2}\beta\right].
\label{eq20}
\end{equation}
Then, we multiply (\ref{eq20}) by $\tilde U(-p)$ and integrate to obtain
\begin{equation}
\alpha=\frac{i\epsilon}{2\pi}\left[\alpha\int_{-\infty}^\infty dp\,\frac{|\tilde
U(p)|^2}{E-p^2}-\beta\int_{-\infty}^\infty dp\,\frac{[\tilde U(p)]^2}{E-p^2}
\right],
\label{eq21}
\end{equation}
and we multiply (\ref{eq20}) by  $\tilde U(p)$ and integrate to obtain
\begin{equation}
\beta=\frac{i\epsilon}{2\pi}\left[\alpha\int_{-\infty}^\infty dp\,\frac{[\tilde
U(p)]^2}{E-p^2}-\beta\int_{-\infty}^\infty dp\,\frac{|\tilde U(p)|^2}{E-p^2}
\right].
\label{eq22}
\end{equation}

These two equations have the form
\begin{equation}
\alpha=\frac{i\epsilon}{2\pi}\alpha I_2-\frac{i\epsilon}{2\pi}\beta I_1,
\qquad \beta=\frac{i\epsilon}{2\pi}\alpha I_1-\frac{i\epsilon}{2\pi}\beta I_2,
\label{eq23}
\end{equation}
where
\begin{equation}
I_1=\int_{-\infty}^\infty dp\,\frac{[\tilde U(p)]^2}{E-p^2},\qquad
I_2=\int_{-\infty}^\infty dp\,\frac{|\tilde U(p)|^2}{E-p^2}.
\label{eq24}
\end{equation}
Note that $\tilde U(p)$ must be complex [otherwise (\ref{eq23}) reduces to
triviality $\alpha=\beta=0$]. It is necessary to assume here that $I_1$ and
$I_2$ exist. This requires that $\tilde U(p)$ vanish at $p^2=E$. Thus, the
eigenvalues of the Hamiltonian must be zeros of $\tilde U(p)$. (Other solutions
for which $E$ is complex correspond to scattering states. As stated earlier, in
this paper we consider only bound states.)

Because the simultaneous linear equations (\ref{eq23}) are homogeneous a
solution for $\alpha$ and $\beta$ exists only if the determinant of the
coefficients vanishes. This requirement gives a secular equation that determines
the energy $E$:
\begin{equation}
1=\frac{\epsilon^2}{4\pi^2}\left(I_1^2-I_2^2\right),
\label{eq25}
\end{equation}
which is the analog of (\ref{eq13}) for the case of a Hermitian separable
potential.

This secular equation has several important properties and consequences. First,
like the secular equation for all $\cP\cT$-symmetric Hamiltonians, it is real
if $E$ is real \cite{R6}. Note that $I_2$ in (\ref{eq24}) is manifestly real. To
see that $I_1$ in (\ref{eq24}) is real, observe that because $U(x)$ is real,
$\tilde U(p)$ is $\cP\cT$ symmetric; that is, $\tilde U^*(p)=\tilde U(-p)$.
Thus, the change of variable $p\to-p$ establishes the reality of $I_1$. As a
consequence, the roots $E$ of this equation are either real or come in
complex-conjugate pairs. Second, (\ref{eq25}) in conjunction with either of the
two equations in (\ref{eq23}) implies that
\begin{equation}
|\alpha|=|\beta|.
\label{eq26}
\end{equation}
Third, observe that $E$ is a function of the {\it square} of the coupling
constant $\epsilon$ so it is independent of the sign of $\epsilon$. This feature
is typical of $\cP\cT$-symmetric Hamiltonians such as $H=p^2+x^2+i\epsilon x$
and $H=p^2+x^2+\epsilon ix^3$, which have imaginary potentials. Fourth, there is
no solution to (\ref{eq25}) if $E$ is negative. This is because when $E<0$ the
triangle inequality implies that $I_2^2\geq I_1^2$, and thus the sign of the
right side of (\ref{eq25}) is negative for any choice of $\epsilon$. Thus, if
there is a bound state, its energy must be positive.

\begin{figure}[t!]
\begin{center}
\includegraphics[scale=0.68, viewport=0 0 360 241]{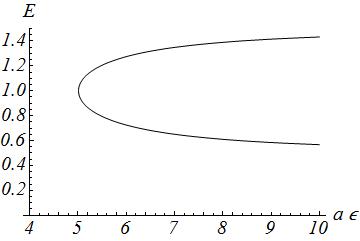}
\end{center}
\caption{The bound-state energy $E$ in (\ref{eq31}) plotted as a function of $a
\epsilon$. The region of unbroken $\cP\cT$ symmetry is $a\epsilon\geq\sqrt{8\pi}
=5.013\ldots$. In this region the Hamiltonian has only one bound state, but
there are two possible allowed values for the energy of this bound-state. As one
can see from (\ref{eq31}), the two allowed values of the energy approach the
asymptotic limits $\frac{3}{2}$ and $\frac{1}{2}$ as $a\epsilon\to\infty$.}
\label{f1}
\end{figure}

To have a bound state for positive energy $E$, the function $\tilde U(p)$ must
vanish at $p^2=E$. As we did in the Hermitian case discussed in Sec.~\ref{s2},
we construct a simple model for which there is just one real root of $\tilde U
(p)$ at $E=p^2$:
\begin{equation}
\tilde U(p)=e^{-p^2/2}(E-p^2)(1+iap),
\label{eq27}
\end{equation}
where $a$ is a {\it real} constant so that $U(x)$ is a real function, namely,
\begin{equation}
U(x)=\frac{1}{\sqrt{2\pi}}e^{-x^2/2}[E-1+(E-3)ax+x^2+ax^3].
\label{eq27.5}
\end{equation}
[Note that if we had chosen this $U(x)$ first, it would not have been obvious
that $\tilde U(p)$ had a zero.] Then
\begin{eqnarray}
I_1&=&\sqrt{\pi}\left(E-\frac{1}{2}-\frac{1}{2}a^2E+\frac{3}{4}a^2\right),
\nonumber\\
I_2&=&\sqrt{\pi}\left(E-\frac{1}{2}+\frac{1}{2}a^2E-\frac{3}{4}a^2\right).
\label{eq28}
\end{eqnarray}
Hence,
\begin{equation}
I_1-I_2 = \frac{1}{2}a^2\sqrt{\pi}(3-2E),\qquad I_1+I_2=\sqrt{\pi}(2E-1).
\label{eq29}
\end{equation}
and the secular equation (\ref{eq25}) reduces to the quadratic equation
\begin{equation}
\frac{8\pi}{a^2\epsilon^2}+(2E-1)(2E-3)=0,
\label{eq30}
\end{equation}
whose roots are
\begin{equation}
E=1\pm\frac{1}{2}\sqrt{1-\frac{8\pi}{a^2\epsilon^2}}.
\label{eq31}
\end{equation}
This equation gives a condition on the coupling constant $\epsilon$ for the
reality of the bound-state energy $E$:
\begin{equation}
\epsilon\geq\frac{\sqrt{8\pi}}{a}.
\label{eq32}
\end{equation}
The interpretation of this inequality is straightforward. Even though $U(x)\to0$
for large $|x|$ it is still possible for there to be a bound state if $\epsilon$
is large enough. However, when $\epsilon$ is smaller than a critical value, the
bound state disappears (its energy becomes complex). This corresponds to going
from a $\cP\cT$-unbroken region to a $\cP\cT$-broken region. A plot of $E$ as a
function of $a\epsilon$ is given in Fig.~\ref{f1}. Note that the Hamiltonian
has only {\it one} bound state; however, there are two allowed values for the
energy of this bound state.

Finally, we must verify that apart from a constant multiplicative phase [which
is arbitrary because the Schr\"odinger equation (\ref{eq17}) is linear] the
eigenfunction $\psi(x)$ is $\cP\cT$ symmetric. To do so, we calculate $\tilde
\psi(p)$ and verify that apart from a constant multiplicative phase the function
$\tilde \psi(p)$ is a real function of $p$. To obtain $\tilde\psi(p)$ we
substitute (\ref{eq27}) into (\ref{eq20}) and obtain
\begin{eqnarray}
\tilde\psi(p)&=&\frac{\epsilon}{2\pi}e^{-p^2/2}[i(\alpha-\beta)-ap
(\alpha+\beta)]\nonumber\\
&=&\frac{\epsilon}{2\pi}(\alpha+\beta) e^{-p^2/2}\left(i\frac{\alpha-\beta}
{\alpha+\beta}-ap\right).\nonumber\\
\label{eq33}
\end{eqnarray}
Next, we add the two equations in (\ref{eq23}) and immediately obtain the result
that the ratio $(\alpha-\beta)/(\alpha+\beta)=-2\pi i/(I_1+I_2)$ is imaginary.
This verifies that up to an arbitrary multiplicative phase, $\psi(x)$ is $\cP
\cT$ symmetric; it becomes $\cP\cT$ symmetric when $\alpha+\beta$ is chosen to
be real, so that $\alpha=\beta^*$. If the coupling constant $\epsilon$ lies
below the critical point $\epsilon<2\pi\sqrt{2}/a$, then $E$ becomes complex;
consequently, $I_1$ and $I_2$ are not real and $\psi(x)$ is not $\cP\cT$
symmetric.

\section{Separated $\cP\cT$-symmetric Potential Having Two Bound States}
\label{s4}

Although it becomes technically complicated, it is straightforward to generalize
the discussion of Sec.~\ref{s3} to the case in which the $\cP\cT$-symmetric
separated potential has more than one bound state. In this section we consider
the case in which there are two bound states of energy $E_1$ and $E_2$ for the
Schr\"odinger equation (\ref{eq17}).

The bound-state eigenfunctions are labeled $\psi_k(x)$ ($k=1,2$), and in
momentum space they have the same general form as in (\ref{eq20}):
\begin{equation}
\tilde\psi_k(p)=\frac{i\epsilon}{2\pi}\left[\frac{\tilde U(p)}{E_k-p^2}\alpha_k
-\frac{\tilde U(-p)}{E_k-p^2}\beta_k\right]\quad(k=1,2),
\label{eq34}
\end{equation}
where
\begin{equation}
\alpha_k=\int_{-\infty}^\infty dq\,\tilde U(-q)\tilde\psi_k(q),\qquad
\beta_k=\int_{-\infty}^\infty dq\,\tilde U(q)\tilde\psi_k(q)\quad(k=1,2).
\label{eq35}
\end{equation}

Equations (\ref{eq21}) and (\ref{eq22}) generalize to
\begin{eqnarray}
\alpha_k&=&\frac{i\epsilon}{2\pi}\left[\alpha_k\int_{-\infty}^\infty dp\,
\frac{|\tilde U(p)|^2}{E_k-p^2}-\beta_k\int_{-\infty}^\infty dp\,\frac{[\tilde
U(p)]^2}{E_k-p^2}\right]\quad(k=1,2),\nonumber\\
\beta_k&=&\frac{i\epsilon}{2\pi}\left[\alpha_k\int_{-\infty}^\infty dp\,
\frac{[\tilde U(p)]^2}{E_k-p^2}-\beta_k\int_{-\infty}^\infty dp\,\frac{|\tilde
U(p)|^2}{E_k-p^2}\right]\quad(k=1,2).
\label{eq36}
\end{eqnarray}
We then extend (\ref{eq27}) so that $\tilde U(p)$ has two real zeros instead
of one:
\begin{equation}
\tilde U(p)=e^{-p^2/2}(E_1-p^2)(E_2-p^2)(1+iap).
\label{eq37}
\end{equation}
The secular equations satisfied by the energies $E_1$ and $E_2$ are a rather
complicated generalization of (\ref{eq30}):
\begin{eqnarray}
\frac{128\pi}{a^2\epsilon^2}+P(E_1,E_2)Q(E_1,E_2)&=&0,\nonumber\\
\frac{128\pi}{a^2\epsilon^2}+P(E_2,E_1)Q(E_2,E_1)&=&0,
\label{eq38}
\end{eqnarray}
where
\begin{eqnarray}
P(E_1,E_2)&=&-15+6E_1+12E_2-8E_1E_2-4E_2^2+8E_1E_2^2,\nonumber\\
Q(E_1,E_2)&=&-105+30E_1+60E_2-24E_1E_2-12E_2^2+8E_1E_2^2.
\label{eq39}
\end{eqnarray}

\begin{figure}[t!]
\begin{center}
\includegraphics[scale=0.68, viewport=0 0 360 251]{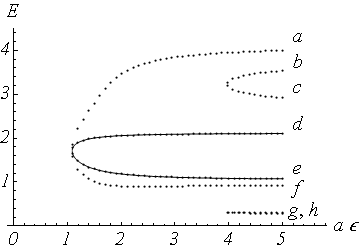}
\end{center}
\caption{Numerical solution to the simultaneous equations (\ref{eq38}) and
(\ref{eq39}). Bound-state energies are graphed as functions of $a\epsilon$. The
region of unbroken $\cP\cT$ symmetry is $a\epsilon\geq 1.09$, and bound states
appear as soon as $a\epsilon$ exceeds this critical value. A second critical
point is at $a\epsilon=3.90$, at which four new solutions appear. There are two
cases: In the first case there is only one bound state whose energy may lie on
the curve $d$ or on the curve $e$. In the second case there are two bound-state
energies, which may lie on the curves $a$ and $f$, or on $b$ and $g$, or on $c$
and $h$. The two curves $g$ and $h$ are too close to be resolved in this figure
and are thus shown in detail in Fig.~\ref{f3}. As $a\epsilon\to\infty$, the
curves $a$ - $h$ approach the asymptotic values $4.081$, $3.742$, $2.725$,
$2.115$, $1.054$, $0.919$, $0.296$, and $0.275$.}
\label{f2}
\end{figure}

\begin{figure}[t!]
\begin{center}
\includegraphics[scale=0.68, viewport=0 0 360 214]{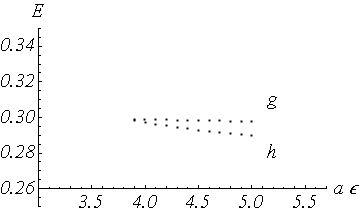}
\end{center}
\caption{Blow-up of the curves $g$ and $h$ from Fig.~\ref{f2}.}
\label{f3}
\end{figure}

By solving simultaneously the equations (\ref{eq38}) and (\ref{eq39}), we obtain
the allowed bound state energies. The numerical solution of these equations is
plotted in Fig.~\ref{f2}. There is a critical point at $a\epsilon=1.09$; below
this value there are no real solutions. Bound states first appear when $a
\epsilon$ exceeds this critical value and there are two cases: (i) There can be
just one bound state if $E_1=E_2$, and in this case there are two possible
values for this energy and these are indicated in Fig.~\ref{f2} by the curves
consisting of connected dots and labeled $d$ and $e$. (ii) There can be two
distinct bound states; these are indicated by the sequences of dots labeled $a$
and $f$. If $E_1$ lies on the curve $a$, then $E_2$ lies on $f$, and {\it vice
versa}. It is interesting that there is a second critical point at $a\epsilon=
3.90$ and when $a\epsilon$ exceeds this value four new possible energies appear;
these energies lie on the curves labeled by $b$, $c$, $g$, and $h$. The possible
bound-state energies $E_1$ and $E_2$ are only allowed to lie on the pairs of
curves $a$ and $f$, or $b$ and $g$, or $c$ and $h$. As $a\epsilon\to\infty$, the
energies on the curves $a$ - $h$ approach the asymptotic values $4.081$,
$3.742$, $2.725$, $2.115$, $1.054$, $0.919$, $0.296$, and $0.275$. The energies
on the curves $g$ and $h$ are too close together to be resolved in
Fig.~\ref{f2}, so we have included a separate figure Fig.~\ref{f3} to show their
dependence on $a\epsilon$.

\section{Comments and discussion}
\label{s5}

We have shown in this paper that it is easy to construct non-Hermitian $\cP
\cT$-symmetric separable potentials and, even though such potentials are
nonlocal, it is still possible to find the secular equation that determines the
bound-state energies. As is the case for any $\cP\cT$-symmetric potential, the
secular equation is real. If one solves the secular equation, one finds a result
that is typical of $\cP\cT$-symmetric theories; namely, that the coupling
constant lies in one of two regions, which are separated by a critical value: On
one side of the critical value (the unbroken region of unbroken $\cP\cT$
symmetry) the energies are real, but on the other side of the critical point
(the region of broken $\cP\cT$ symmetry) the energies are complex. This
$\cP\cT$ phase transition has been observed experimentally in $\cP\cT$-symmetric
optical models \cite{R7,R8}.

Based on the structure and behavior of the models we have constructed in this
paper, it is evident that we can construct $\cP\cT$-symmetric separable
potentials for which there are as many bound states as we wish, and in the
unbroken $\cP\cT$-symmetric region all of the bound states will have positive
energies.

For the models discussed in this paper the potentials vanish exponentially
rapidly as $|x|\to\infty$. Thus, for large $|x|$ the Hamiltonian becomes the
free Hamiltonian $H_0=p^2$, whose solutions are plane waves. Thus, in addition
to bound states, there will be scattering states for all positive energies. The
energy of a scattering state will be complex, with the sign of the imaginary
part of the energy being associated with incoming- or outgoing-wave boundary
conditions. The models we have studied in this paper are interesting because the
point spectrum of bound states is embedded in the continuum of scattering states
\cite{R9,R10,R11}.

\begin{acknowledgments}
CMB is supported by a grant from the U.S.~Department of Energy.
\end{acknowledgments}

\end{document}